\documentclass[prl,reprint,twocolumn,superscriptaddress,showpacs]{revtex4-1}

\pdfoutput=1

\usepackage{graphicx}
\usepackage{amsmath,amssymb}

\usepackage{ulem}
\usepackage{color}
\usepackage{amssymb}

\def\diff{\mathrm d}

\newcommand{\Tc}{\ensuremath{T_\mathrm{c}}}
\newcommand{\jeff}{\ensuremath{{j_\mathrm{eff}}}}
\newcommand{\Dtri}{\ensuremath{\Delta_\mathrm{tri}}}
\newcommand{\JH}{\ensuremath{J_\mathrm{H}}}

\begin{document}
\title{
Phase diagram of pyrochlore iridates: all-in--all-out magnetic ordering and non-Fermi liquid properties
}
\author{Hiroshi Shinaoka} 
\affiliation{Theoretische Physik, ETH Z\"{u}rich, 8093 Z\"{u}rich, Switzerland}
\affiliation{Department of Physics, University of Fribourg, 1700 Fribourg, Switzerland}
\author{Shintaro Hoshino}
\affiliation{Department of Basic Science, The University of Tokyo, Meguro 153-8902, Japan}
\affiliation{Department of Physics, University of Fribourg, 1700 Fribourg, Switzerland}
\author{Matthias Troyer}
\affiliation{Theoretische Physik, ETH Z\"{u}rich, 8093 Z\"{u}rich, Switzerland}
\author{Philipp Werner}
\affiliation{Department of Physics, University of Fribourg, 1700 Fribourg, Switzerland}
\date{\today}

\begin{abstract}
We study the prototype $5d$ pyrochlore iridate Y$_2$Ir$_2$O$_7$ from first principles using the local density approximation and dynamical mean-field theory (LDA+DMFT).
We map out the phase diagram in the space of temperature, onsite Coulomb repulsion $U$, and filling.
Consistent with experiments, we find that an all-in--all-out ordered insulating phase is stable for realistic values of $U$. 
The trigonal crystal field enhances the hybridization between the $\jeff$ =1/2 and $\jeff=3/2$ states, and strong inter-band correlations are induced by the Coulomb interaction, which indicates that a three-band description is important.
We demonstrate a substantial band narrowing in the paramagnetic metallic phase and non-Fermi liquid behavior in the electron/hole doped system originating from long-lived quasi-spin moments induced by nearly flat bands. 
\end{abstract}


\maketitle

The competition and cooperation between spin-orbit coupling (SOC) and electron correlations induces novel phenomena
in 4$d$ and 5$d$ transition metal oxides such as spin-orbit-assisted Mott insulators, topological phases, and spin liquids~\cite{WitczakKrempa:2014hz}.
The pyrochlore iridates $A_2$Ir$_2$O$_7$ ($A$=Pr, Nd, Y, \textit{etc}.) 
are an ideal system to study these phenomena because their magnetic and electronic states can be tuned by chemical substitution, pressure, and temperature ($T$).
Furthermore, intriguing phenomena such as correlated topological phases have been theoretically predicted on their geometrically frustrated crystal structure~\cite{WitczakKrempa:2014hz}.

\begin{figure}
 \centering
 \includegraphics[width=.5\textwidth,clip]{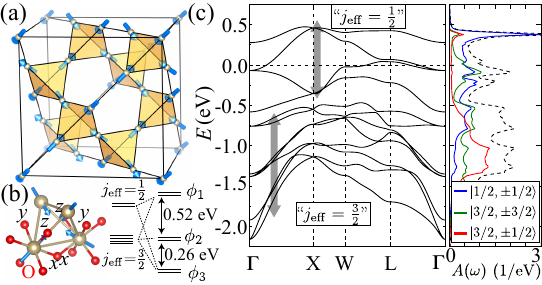}
 \caption{(color online). 
(a) Pyrochlore lattice formed by Ir atoms with arrows representing spin moments in the all-in--all-out magnetic structure.
(b) fcc unit cell with the local coordinate axes and the energy diagram under SOC ($\zeta$) and the trigonal crystal field ($\Delta_\mathrm{tri}$).
(c) LDA band structure together with the density of states projected on the $\jeff$ basis. The broken line shows the total density of states.
The ``$\jeff$=1/2" and ``$\jeff$=3/2" bands consists mainly of the $\jeff$=1/2 and $\jeff$=3/2 manifolds, but are substantially hybridized under the trigonal CF (see the text).
}
 \label{fig:sys}
\end{figure}

In 2001, it was reported that these compounds show a crossover from metal to insulator with decreasing $A^{3+}$ ionic radii at high $T$~\cite{Yanagishima:2001ej} and a
magnetic anomaly was found at low $T$ for small $A^{3+}$ ionic radii~\cite{Taira:2001uz}.
For the metallic compound $A$=Pr,
experiments revealed spin-liquid behavior~\cite{Nakatsuji:2006fc,Tokiwa:2014ik} and an unconventional anomalous Hall effect~\cite{Machida:2007dc}.
On the other hand, 
the Ir magnetic ordering has not been determined for a decade due to the strong neutron absorption by Ir and
large magnetic contributions from rare-earth $f$ moments on $A^{3+}$.
The magnetic order has only recently been identified as a noncollinear all-in--all-out order [see Fig.~\ref{fig:sys}(a)]~\cite{Tomiyasu:2012fx,Sagayama:2013fn,Disseler:2014ie}.

Among the insulating compounds, Y$_2$Ir$_2$O$_7$ has the highest magnetic transition temperature and no $f$ moments.
This makes this compound a prototype system for studying strong electron correlations among 5$d$ electrons.
A pioneering local density approximation (LDA)+$U$ study for this compound showed that the all-in--all-out order is indeed stable at large on-site repulsion $U$~\cite{Wan:2011hi}.
It also proposed a topological Weyl semimetal as the ground state of some compounds in this series.
This stimulated further theoretical studies on the topological nature and unconventional quantum criticality of 5$d$ electrons on the pyrochlore lattice~\cite{Yang:2010dv,Anonymous:2011df,Chen:2012eka,WitczakKrempa:2012di,Moon:2012ha,Go:2012fb,Lee:2013fv,Savary:2014wf,WitczakKrempa:2014hz,Ruegg:2015vu}.

In pyrochlore iridates, the Ir atoms form a 
frustrated pyrochlore lattice, a corner-sharing network of tetrahedra [see Fig.~\ref{fig:sys}(a) and Ref.~\onlinecite{Gardner:2010fu}].
The so-called $\jeff$=1/2 picture was originally proposed for the insulating quasi-2D compound Sr$_2$IrO$_4$ with the same electron configuration $5d^5$~\cite{Kim:2008gi,Kim:2009dg}.
The SOC splits the $t_\mathrm{2g}$ manifold into a fully occupied $\jeff$=3/2 quartet and a half-filled $\jeff$=1/2 doublet ($\hat{j}_\mathrm{eff}\equiv \hat{S}-\hat{L}$ is the effective total angular momentum, $\hat{S}$ and $\hat{L}$ the spin and orbital momenta).
This picture was subsequently confirmed by LDA+dynamical mean-field theory (DMFT) studies~\cite{Martins:2011kj,Arita:2012ei,Zhang:2013iy}.
The $\jeff$=1/2 and 3/2 manifolds hybridize either under a tetragonal crystal field (CF), as in the case of Sr$_2$IrO$_4$, or under a trigonal CF
because they do not commute with $\hat{j}_\mathrm{eff}^2$.
For pyrochlore iridates, a recent quantum chemistry calculation found that the trigonal CF is comparable to the SOC~\cite{Hozoi:2014ft}.
Moreover, the $t_\mathrm{2g}$ band width in the LDA band structure ($\simeq$ 2 eV~\cite{Singh:2008dk}) is even smaller than the typical value of $U$ for Ir$^{4+}$ ($\simeq$ 3 eV~\cite{Yamaji:2014be}).
These effects will enhance the hybridization between the two $\jeff$ manifolds beyond the single-particle level, raising questions about the 
validity of the $\jeff$=1/2 single-band picture.
Another interesting issue is the role of the geometrical frustration,
which suppresses long-range magnetic ordering and induces a novel type of magnetism~\cite{Gardner:2010fu}.
To address these questions, it is important to perform a first-principles study 
of the three-band system which takes into account SOC, electron correlations, itinerancy, and the crystal structure.

In this Letter, we present state-of-the-art relativistic LDA+DMFT calculations for the prototypical compound Y$_2$Ir$_2$O$_7$.
We map out the phase diagram in $T$ and on-site Coulomb repulsion $U$ and investigate the properties of the correlated 5$d$ electrons in this family.
First, we construct maximally localized Wannier functions for the $t_\mathrm{2g}$ manifold~\footnote{For technical details, refer to Ref.~\onlinecite{PhysRevB.88.174422}.} using an LDA exchange-correlation functional~\cite{Ceperley:1980zz,Perdew:1981dv}.
We use the code QMAS (Quantum MAterials Simulator)~\cite{qmas}, which is based on the projector augmented wave method~\cite{Blochl:1994dx}, and the two-component formalism~\cite{Kosugi:2011gm,Oda:1998jc}.
The experimental crystal structure at 290 K is taken from Ref.~\onlinecite{Anonymous:lj_4oZcT}.
In our DMFT calculations, electron correlation effects are taken into account by introducing the Slater-Kanamori interaction
\begin{eqnarray}
&&H_\text{int}=\frac{1}{2}\sum_{\alpha\beta\alpha^\prime\beta^\prime\sigma\sigma^\prime}U_{\alpha\beta\alpha^\prime\beta^\prime}c^\dagger_{i\alpha\sigma}c^\dagger_{i\beta\sigma^\prime}c_{i\beta^\prime\sigma^\prime}c_{i\alpha^\prime\sigma}
\end{eqnarray}
in the standard parameterization 
$U_{\alpha\alpha\alpha\alpha}=U$, $U_{\alpha\beta\alpha\beta}=U-2J_\mathrm{H}$, $U_{\alpha\beta\beta\alpha}=U_{\alpha\alpha\beta\beta}=J_\mathrm{H}$ ($\alpha\neq \beta$), with 
$\alpha$ ($\beta$) and $\sigma$ ($\sigma^\prime$) being orbital and spin indices, respectively.
$U$ and $J_\mathrm{H}$ are the on-site repulsion and the Hund's coupling, respectively.
We choose $J_\mathrm{H}/U=0.1$, which is motivated by a first-principles estimate for the related compound Na$_2$IrO$_3$ ($U$=2.72 eV, $J_\mathrm{H}=0.23$ eV)~\cite{Yamaji:2014be}.
Within DMFT, one has to solve a three-orbital quantum impurity problem with off-diagonal and complex hybridization functions.
We employ a numerically exact continuous-time quantum Monte Carlo impurity solver based on the hybridization expansion~\cite{Werner:2006ko,Werner:2006iz}.
In previous studies,
the quantum impurity models for $5d$ electrons have been simplified to avoid a severe sign problem, e.g., by omitting off-diagonal hybridization functions and some interaction terms in the $\jeff$ basis~\cite{privatecomm:xidai}.
Since pyrochlore iridates have large inter-band hybridizations,
we solve our impurity problem without such approximations.
Another advantage of this treatment is that we do not necessarily need to assume the quantization axes of spin and orbital.
The sign problem is reduced by rotating the single-particle basis of the hybridization function~\footnote{Refer to Appendix B in Ref.~\onlinecite{Shinaoka:2015hm}}.
Please refer to the Supplemental Material for technical details and some results on the effects of the off-diagonal hybridizations.

Figure~\ref{fig:sys}(c) shows the computed LDA band structure.
The upper half-filled manifold, which is usually identified as the $\jeff$=1/2 manifold, has an overlap with the lower manifold in energy space,
although the bands are separated at each $k$ point.
The $\jeff$=1/2 manifold has four Kramers degenerate bands since a unit cell contains four Ir atoms.
We constructed a tight-binding model based on $t_\mathrm{2g}$-orbital-like maximally localized Wannier functions.
The SOC $\zeta$ and the trigonal crystal field $\Dtri$ are estimated to be $\zeta=0.40$ eV and $\Dtri=0.23$ eV~\footnote{We used the same notation in Ref.~\onlinecite{PhysRevB.88.174422}}.
These values are consistent with an estimate by a quantum chemistry calculation~\cite{Hozoi:2014ft}.
As shown in Fig.~\ref{fig:sys}(b),
the $t_\mathrm{2g}$ manifold splits into three doublets under $\zeta$ and $\Dtri$.
The wavefunction of the highest doublet $\phi_1$ is given by $\phi_{1\pm} = -0.977|1/2,\pm 1/2\rangle - 0.212|3/2,\pm 1/2\rangle$ in the $\jeff$ basis $|\jeff, j_\mathrm{eff}^{111}\rangle$.
We denote by $\hat{j}_\mathrm{eff}^{111}$ the effective angular momentum along the local [111] axis [see Fig.~\ref{fig:sys}(a)].
The $|1/2,\pm 1/2\rangle$ states have about 71\% reduced spin and orbital moments compared to the ideal atomic values 1/3$\mu_\mathrm{B}$ and 2/3$\mu_\mathrm{B}$, because the Wannier functions have substantial weight on neighboring oxygen atoms.
On the other hand, the magnetic moments are enhanced by the hybridization between the $\jeff$=1/2 and $\jeff$=3/2 manifolds by $\Dtri$.
As a result, the doublet $\phi_1$ has spin and orbital moments of $0.490\mu_\mathrm{B}$ and $0.598\mu_\mathrm{B}$.
To illustrate the effects of itinerancy, we plot the density of states projected on the $\jeff$ basis.
The contributions of $|1/2,\pm 1/2\rangle$ and $|3/2, \pm 3/2\rangle$, which are not mixed by $\Dtri$, have comparable weight near the Fermi level,
which indicates that the inter-atomic hybridization also plays a substantial role.
\begin{figure}
 \begin{tabular}{c}
 \begin{minipage}{0.6\hsize}
     \centering
     \includegraphics[width=\textwidth,clip]{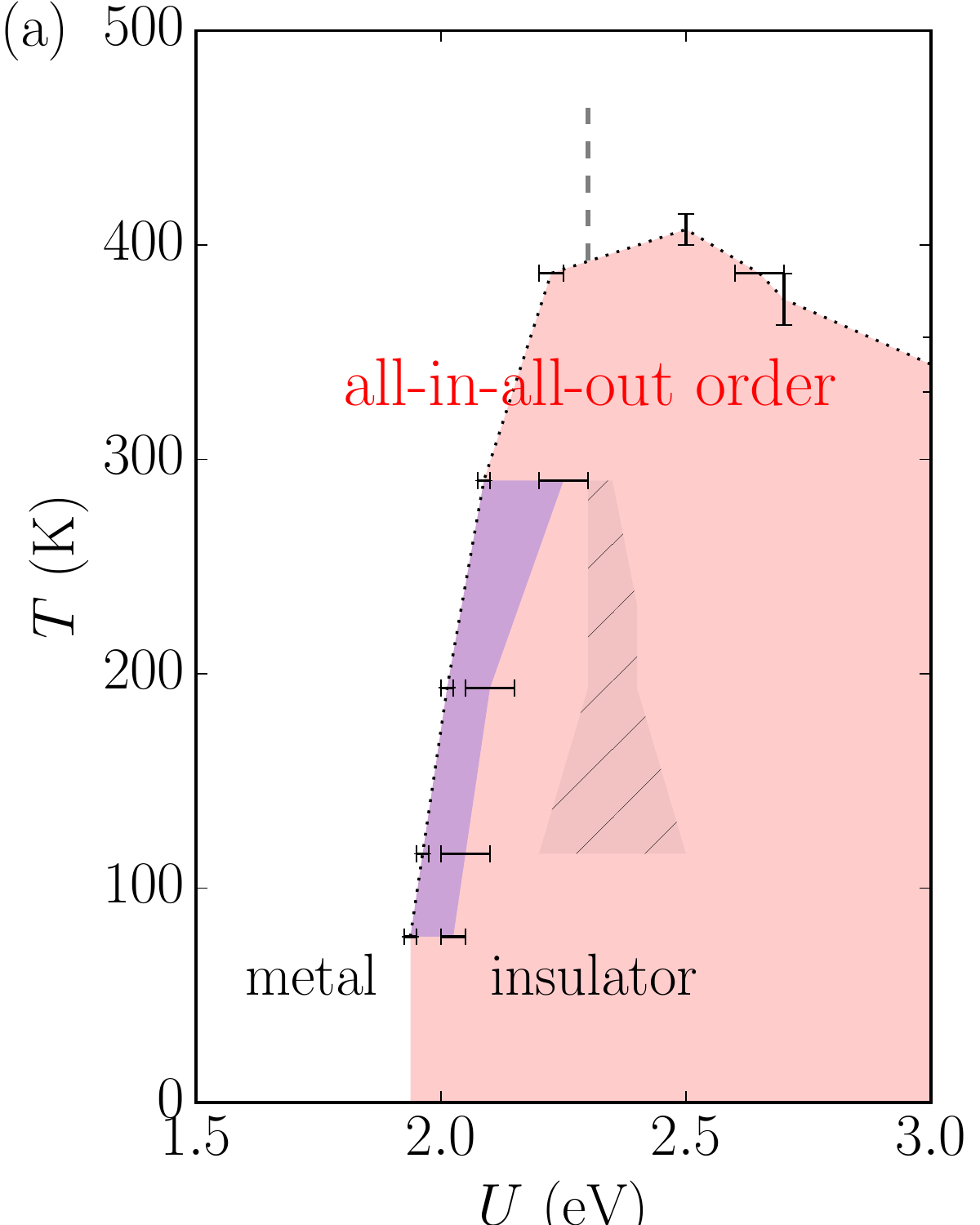}
 \end{minipage}
 \begin{minipage}{0.4\hsize}
 \includegraphics[width=\textwidth,clip]{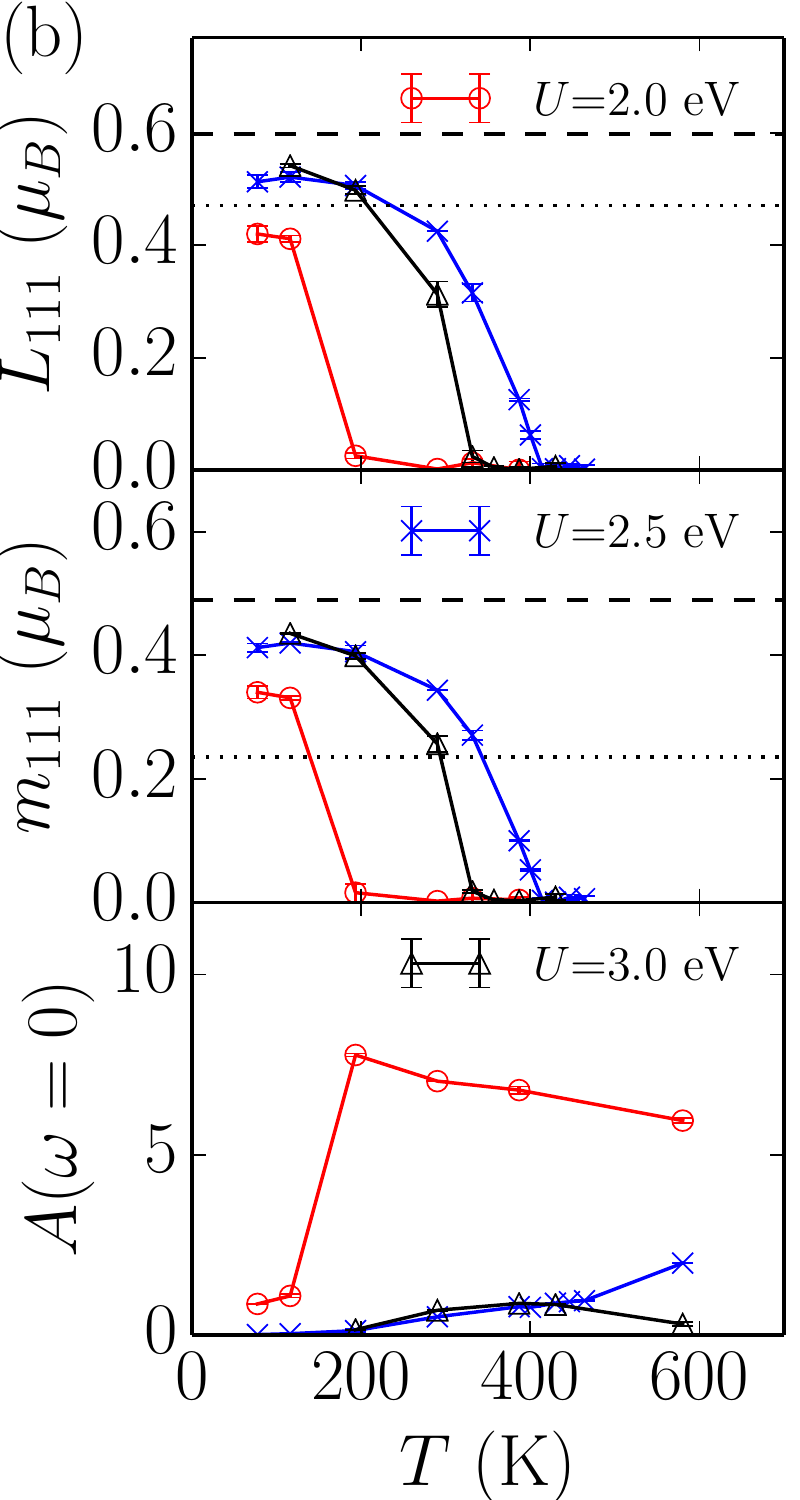}
 \end{minipage}
 \end{tabular}
 \caption{(color online). 
 (a) $U$-$T$ phase diagram at half filling.
 There is a first-order transition between the magnetic insulator and the paramagnetic metal at low $T$ and small $U$.
 The blue shaded region denotes the hysteresis region associated with this transition.
 The error bars reflect the uncertainty caused by the finite number of parameter values considered.
 The metal-insulator crossover in the high-$T$ paramagnetic phase is shown by a broken line.
 The hashed region represents the first-order Mott transition and its hysteresis region in paramagnetic DMFT calculations.
 (b) temperature dependence of the angular and magnetic moment along the local [111] axis and the spectral weight at $\omega=0$ for half filling.
 The moment values of the $\jeff$=1/2 and $\phi_1$ doublets are shown by dotted and broken lines, respectively (see text).
  }
 \label{fig:pd}
\end{figure}

Next, we discuss the $U$-$T$ phase diagram obtained by the DMFT calculations [Fig.~\ref{fig:pd}(a)].
There is a dome-shaped all-in--all-out magnetically ordered phase at large $U$.
The transition temperature $\Tc$ rises up to values about three times higher than the experimental $\Tc^\text{exp}\lesssim 150$ K.
This may be due to the neglect of spatial fluctuations in the DMFT approximation.
We show the $T$-dependent spin and orbital moments along the local [111] axis ($m_{111}$ and $L_{111}$) computed at several values of $U$ in Fig.~\ref{fig:pd}(b).
Both order parameters emerge concurrently with the same sign.
The Ir magnetic moment is estimated to be 0.93$\mu_B$ at low $T$ for $U=2.5$ eV, which is larger than an experimental estimate of the upper bound ($0.5\mu_B$)~\cite{Anonymous:lj_4oZcT}.
The ratio $L_{111}/m_{111}\simeq 1.2$ is substantially smaller than $L_{111}/m_{111}=2$ of the $\jeff$=1/2 state.
In contrast, a LDA+DMFT study estimated $L_{111}/m_{111}\simeq 2.2$ for Sr$_2$IrO$_4$~\cite{Zhang:2013iy}.
On the other hand, the spectral weight $A(\omega=0)$ shows a drop at $\Tc$ for $U=2$ eV,
signaling a transition from a paramagnetic metal into an all-in--all-out ordered insulator.
At higher $U$ ($\ge 2.5$ eV),
the spectral function is substantially suppressed even above $\Tc$,
indicating that the system is in a Mott insulating state.
A crossover between a paramagnetic metal and a paramagnetic insulator is located around $U=2.3$ eV at high $T$.
In Fig.~\ref{fig:pd}(a), we also show the first-order Mott transition line obtained by paramagnetic DMFT calculations.
Apparently, the metal-insulator transition in the magnetic DMFT phase diagram is assisted by magnetic ordering.
The insulating compound Y$_2$Ir$_2$O$_7$, which has the highest $\Tc$ in the family, may be located at $U\simeq 2.5$ eV.

At low $T$ and small $U$, we find a first-order transition between the paramagnetic metallic phase and the all-in--all-out ordered insulating phase~(both states are topologically trivial).
For Nd$_2$Ir$_2$O$_7$, it was reported that its magnetic and metal-insulator transition at $\Tc$=33 K is second order~\cite{Matsuhira:2011ku}.
The first-order nature may be an artifact of the LDA+DMFT method.
A previous LDA+$U$ study found a Weyl semimetallic phase on the lower-$U$ side of an all-in--all-out ordered insulating phase~\cite{Wan:2011hi}.
In our DMFT phase diagram, however, the semimetallic phase is taken over by the insulating phase and there is a direct transition between the paramagnetic metal and the magnetic insulator.
This appears to be a strong correlation effect.

\begin{figure}
 \begin{tabular}{c}
 \begin{minipage}{0.8\hsize}
  \centering
  \includegraphics[width=\textwidth,clip,type=pdf,ext=.pdf,read=.pdf]{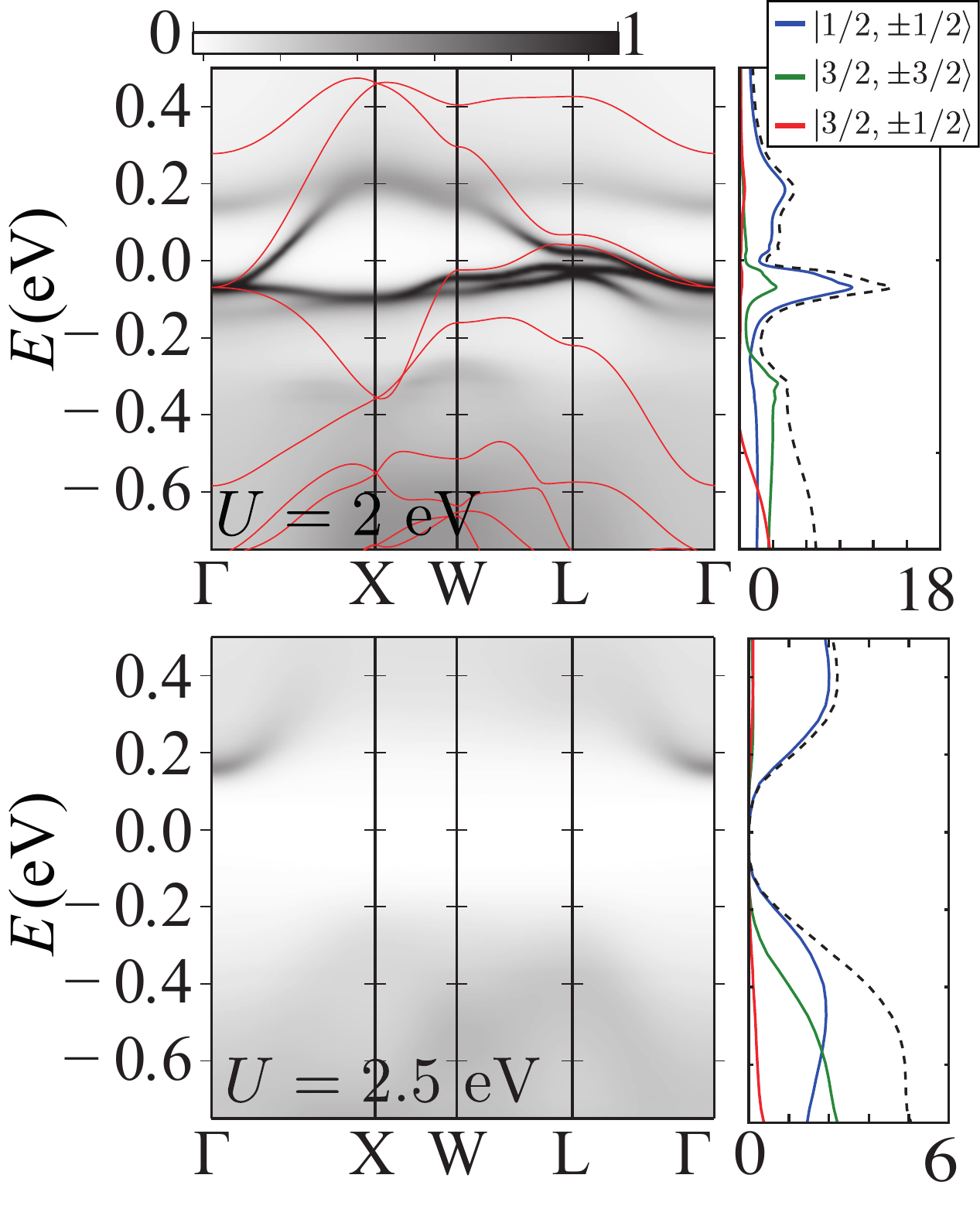}
 \end{minipage}
 \end{tabular}
 \caption{(color online).
 Momentum-resolved spectral function $A(k,\omega)$ at $U=2$ eV (paramagnetic metal) and $U=2.5$ eV (all-in--all-out ordered insulator) at 290 K.
 The LDA band structure is shown by red lines.
 On the right we plot the $k$-integrated spectral function projected on the $\jeff$ basis (the broken line is the total spectral function).
 }
 \label{fig:Ak}
\end{figure}

Figure~\ref{fig:Ak} shows the spectral function $A(k,\omega)$ computed for $T=290$ K.
At $U=2$ eV (paramagnetic metal), the upper manifold near the Fermi level shows a substantial band narrowing from the LDA value of $\simeq 1$ eV down to approximately $0.4$ eV, while the lower manifold is smeared out by correlation effects.
The low-energy states consist mainly of the $\jeff$=1/2 orbitals for $\omega>0$.
A similar purification of the spectral function was found in LDA+DMFT studies of Sr$_2$IrO$_4$.
At $U=2$ eV and $U=2.5$ eV (all-in--all-out ordered insulator), no clear separation is seen between the $j_\text{eff}$=1/2 and $j_\text{eff}$=3/2 manifolds in the total spectral function $A(\omega)$.
\begin{figure}
 \begin{tabular}{cc}
 \begin{minipage}{0.5\hsize}
     \centering
     \includegraphics[width=\textwidth,clip]{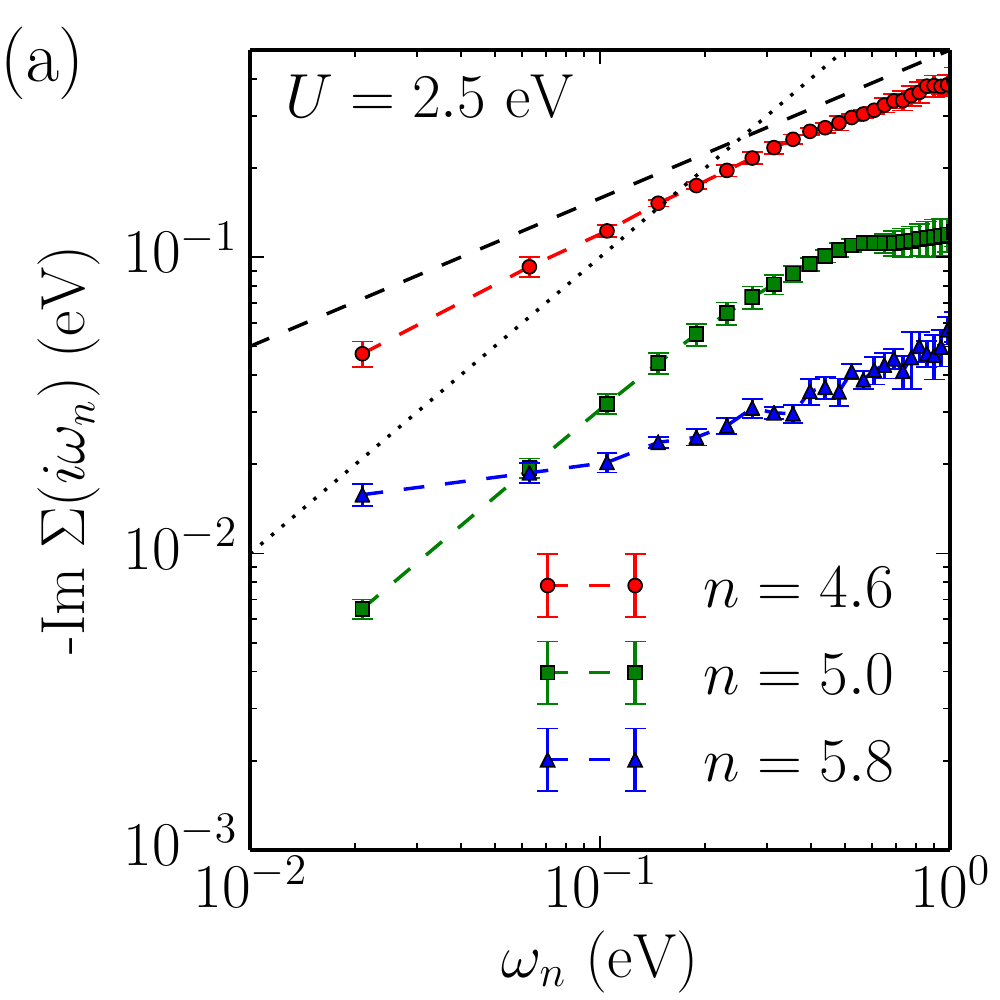}
 \end{minipage}
 &
 \begin{minipage}{0.5\hsize} 
      \includegraphics[width=\textwidth,clip]{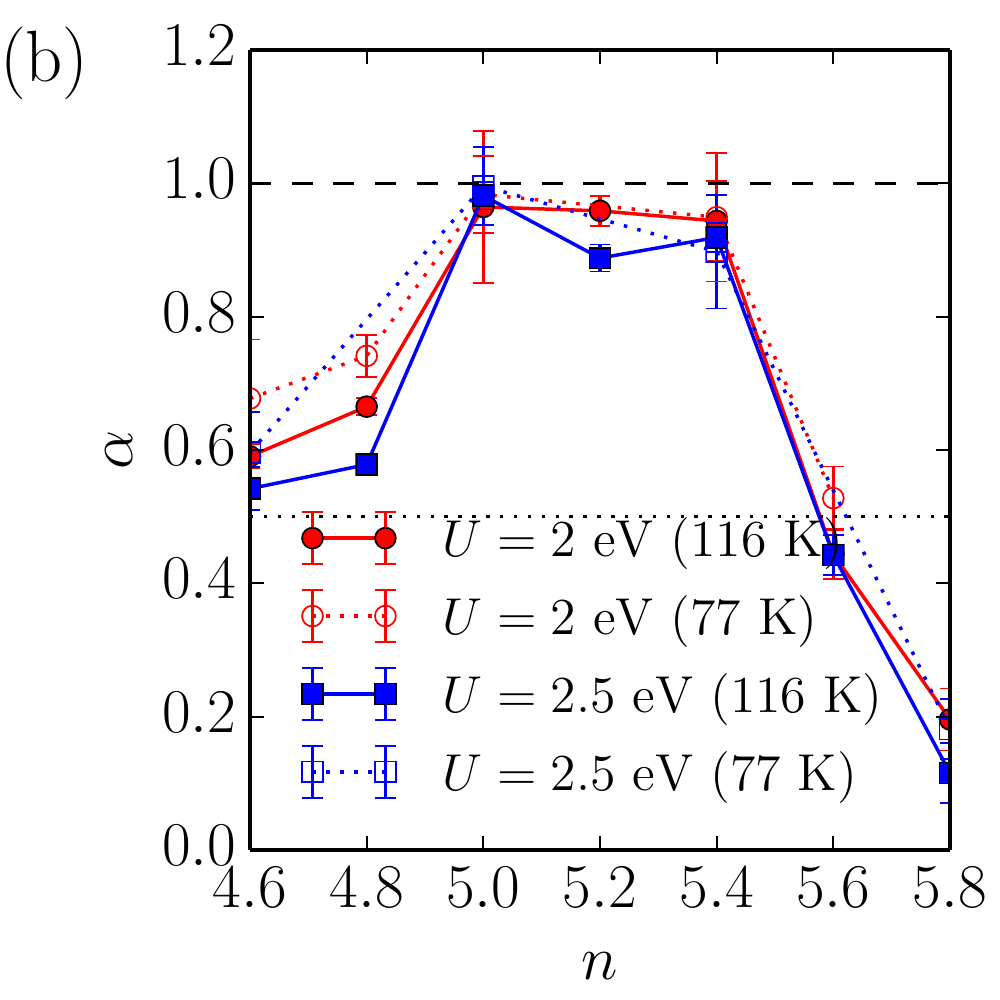}
 \end{minipage}
 \\
 \begin{minipage}{0.5\hsize}
     \centering
     \includegraphics[width=\textwidth,clip]{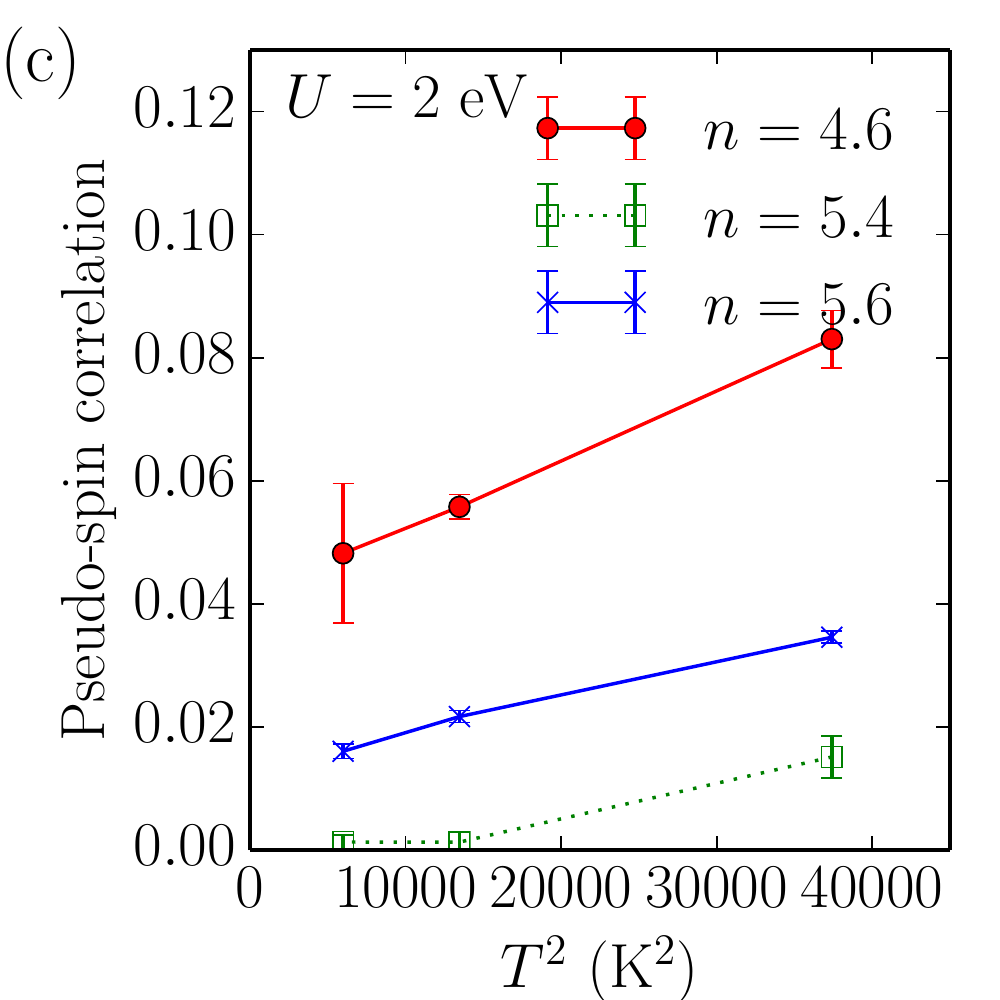}
 \end{minipage}
 &
 \begin{minipage}{0.5\hsize}
     \centering
     \includegraphics[width=\textwidth,clip]{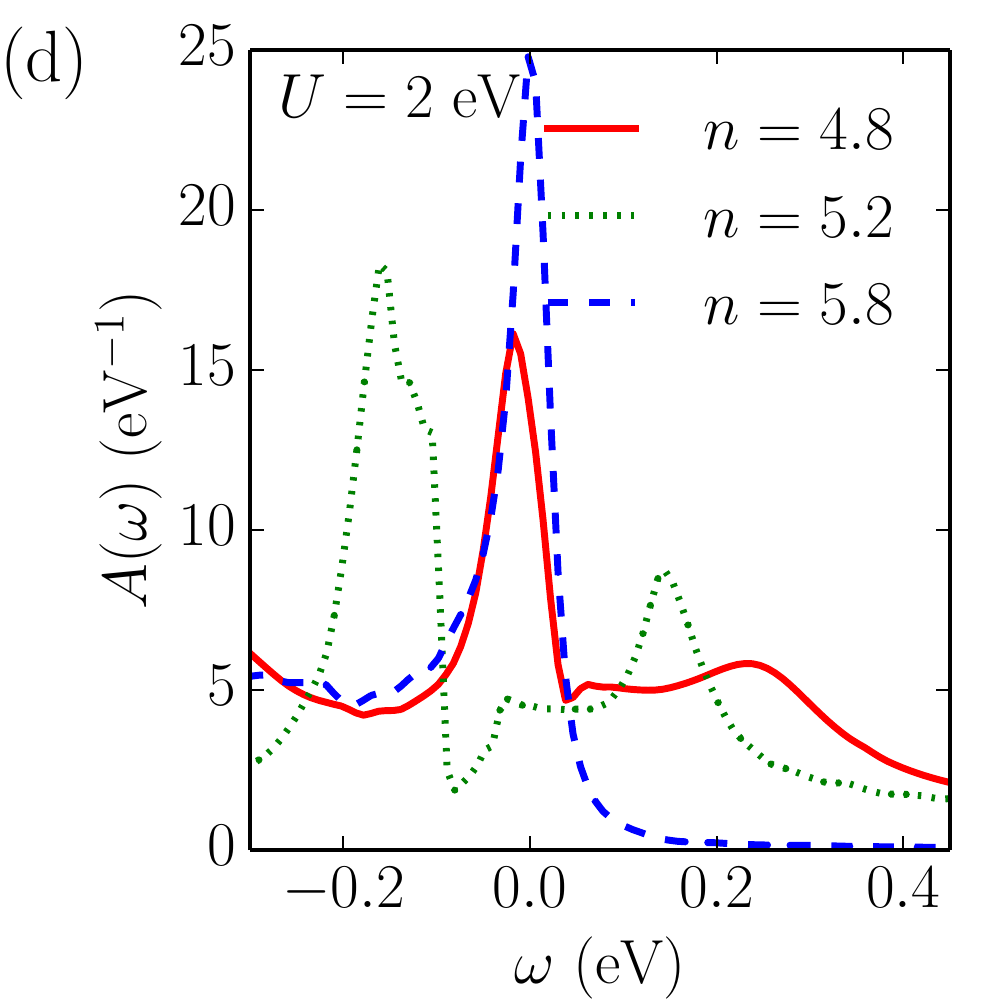}
 \end{minipage}\\
  \begin{minipage}{0.5\hsize}
      \centering
      \includegraphics[width=\textwidth,clip]{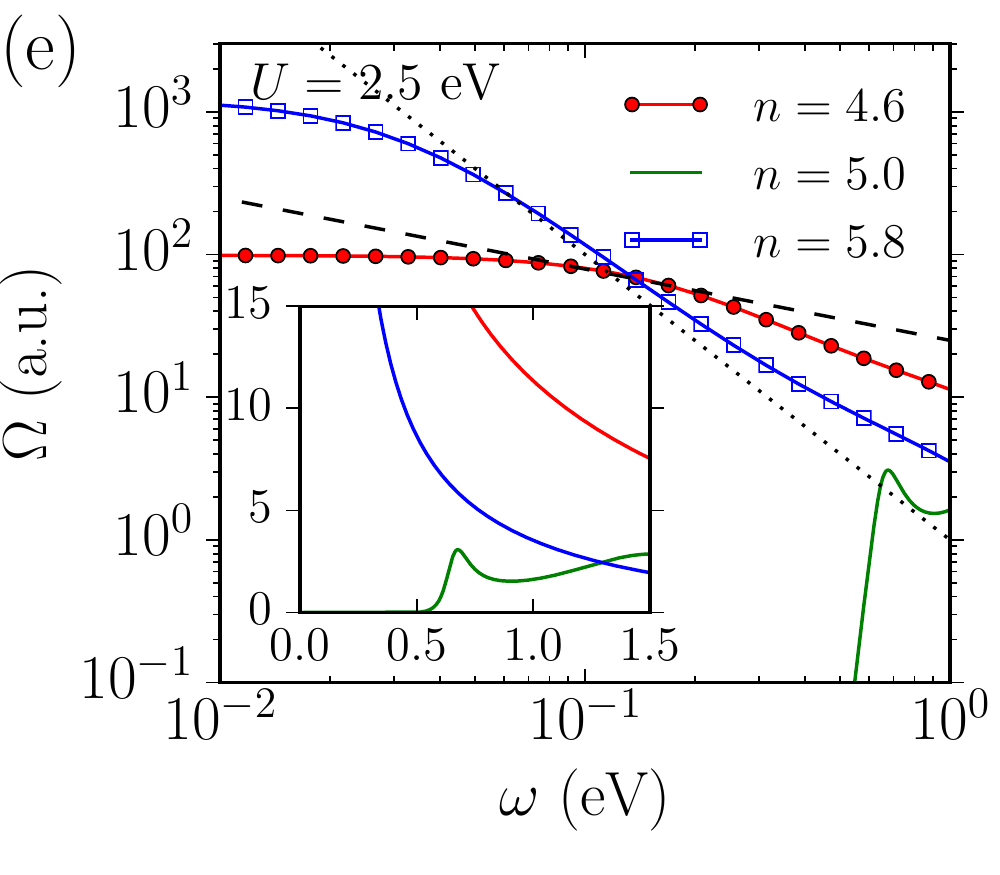}
  \end{minipage}
  &
 \end{tabular}
 \caption{(color online).
 (a) frequency dependence of the imaginary part of the self-energy projected on the $\jeff$=1/2 manifold.
 The dotted and broken lines are $\propto \omega_n$ and $\propto \omega_n^{0.5}$, respectively.
 $n$ is the number of electrons in the $t_\mathrm{2g}$ shell ($n=5$ is the undoped case).
 (b) $n$ dependence of the exponent $\alpha$ [Im $\Sigma(i\omega_n) \propto \omega_n^\alpha$].
 (c) $T$ dependence of the imaginary-time pseudo-spin correlation at $\tau=\beta/2$ (see the text for the definition).
 (d) spectral functions computed at 116 K.
 (e) optical conductivity $\Omega(\omega)$.
 The dotted and broken lines are $\propto \omega^{-2}$ and $\propto \omega^{-0.5}$, respectively.
 }
 \label{fig:NFL}
\end{figure}

We now investigate the effect of doping this insulating solution. The electron filling $n$ is changed from 4.6 to 5.8 ($n=5$ corresponds to the undoped compound).
The all-in--all-out order vanishes rapidly upon electron/hole doping at $n=4.8$ and $n=5.2$, respectively.
We plot Im$\Sigma(i\omega_n)$ in Fig.~\ref{fig:NFL}(a) for $n=4.6$ and $n=5.6$.
In a Fermi liquid, Im$\Sigma(i\omega_n)$ vanishes linearly with $\omega_n$ at low frequencies (i.e, $\propto \omega_n^\alpha$ with $\alpha=1$) when $T$ is sufficiently low.
However, for a rather wide range of frequencies, our data show that Im $\Sigma(i\omega_n)$ vanishes more slowly in the doped insulator.
We estimated the exponent $\alpha$ using the lowest two Matsubara frequencies assuming Im $\Sigma(i\omega_n) = \beta \omega_n^\alpha$.
The result is shown in Fig.~\ref{fig:NFL}(b) for different $n$.
One sees a substantial reduction of $\alpha$ from 1 around $n=4.6$ and $n=5.8$.
This non-Fermi liquid (NFL) or bad metallic behavior persists down to the lowest temperature considered (77 K).

Theoretically, it has been reported that NFL behavior with a reduced power-law exponent in the frequency-dependent self-energy can be induced by the Hund coupling in multi-orbital systems~\cite{Werner:2008fu,deMedici:2011gga,Georges:2013ju}.
In the latter case, $\mathrm{Im} \Sigma(i\omega_n)$ shows a nonzero intersect as $\omega_n\rightarrow 0$ close to half-filling, and a fractional power-law scaling appears at the boundary between the NFL regime and the Fermi liquid regime~\cite{Werner:2008fu}.
This sharp crossover has been coined the ``spin-freezing transition", because the scattering in the NFL state is related to the appearance of long-lived local moments.
To see if the NFL behavior in the pyrochlore iridates has a similar origin, we plot in 
Fig.~\ref{fig:NFL}(c) the imaginary-time correlator $\langle \hat{s}^\mathrm{quasi}(\tau=\beta/2) \hat{s}^\mathrm{quasi}(\tau=0) \rangle$  computed for $U=2$ eV,
where the quasi spin is defined for the highest $\phi_1$ doublet as $\hat{s}^\mathrm{quasi}=\hat{n}_{\phi_{1+}}-\hat{n}_{\phi_{1-}}$.
As can be seen, this correlation function extrapolates to a finite value as $T\rightarrow 0$ for $n=4.6$ and $n=5.8$, indicating the existence of long-lived quasi-spin moments.
However, these long-lived moments are not a consequence of the Hund coupling; they rather 
originate from properties of the band dispersion, i.e. crystal structure,
which we confirmed by reproducing this NFL behavior even with $\JH$ turned off (not shown).

As illustrated in Fig.~\ref{fig:NFL}(d), $A(\omega)$ shows two characteristic peaks at the upper and lower edges of the $\jeff$=1/2 manifold at $n=5.2$ (electron-doped Fermi liquid).
These two peaks come from nearly flat quasi-particle bands in the paramagnetic phase [Fig.~\ref{fig:Ak}(a)], corresponding to peak structures seen in the LDA density of states [Fig.~\ref{fig:sys}©]. 
The NFL behavior appears when the Fermi level hits one of these peaks.
This leads to the formation of local moments and a reduction of the coherence temperature $T^*$.

One might wonder if the NFL state is stable against magnetic ordering other than the all-in--all-out order.
We confirmed that 
the NFL state is stable by performing DMFT calculations without assuming the all-in--all-out magnetic ordering down to 116 K.
This robustness is likely due to the geometrically frustrated crystal structure which prevents long-range ordering.
The NFL state may be observed by optical conductivity measurements (e.g., $\alpha=0.5$ implies an optical conductivity $\Omega(\omega)\propto 1/\sqrt{\omega}$~\cite{Werner:2008fu}),
or by the $T$-dependence of the resistivity.
Figure~\ref{fig:NFL}(e) shows the optical conductivity $\Omega(\omega)$ along [001] computed at $U=2.5$ eV and 116 K.
The $\omega$-dependence deviates substantially from the Fermi-liquid-like behavior $\omega^{-2}$ for the hole-doped case $n=4.6$ around $\omega=1$ eV.
On the other hand, the conductivity of the electron-doped compound $n=5.8$ is closer to a $\omega^{-2}$ behavior, which is likely due to the small self-energy [see Fig.~\ref{fig:NFL}(a)].
It was reported that the undoped metallic Bi$_2$Ir$_2$O$_7$ shows a bad metallic behavior and the transport properties sensitively depend on the magnetic field~\cite{Qi:2012gi}.
That paper discussed the connection between these nontrivial properties and a sharp peak in the LDA density of states near the Fermi level.
Although this compound has a wider band width and is considered to be weakly correlated,
it would be interesting to investigate it with LDA+DMFT. 

In summary, we have mapped out the finite-$T$ phase diagram of the $5d$ pyrochlore iridate Y$_2$Ir$_2$O$_7$ using a state-of-the-art relativistic LDA+DMFT approach.
We showed that the spin and orbital moments are substantially enhanced by the trigonal crystal field and that the $\jeff$=1/2 picture is not valid.
We have also identified a characteristic non-Fermi liquid self-energy which originates from long-lived quasi-spin moments induced by nearly flat bands.
Due to experimental difficulties such as the lack of a natural cleavage plane,
angle-resolved photoemission data of this family have been reported only for Bi$_2$Ir$_2$O$_7$~\cite{Quinn:2013wg}.
However, 
the connection of the non-Fermi liquid behavior to nearly flat bands enables an analysis of the electronic structure by macroscopic measurements.
Although the hole-doped compounds Ca$_x$Y$_{2-x}$Ir$_2$O$_7$ show a bad metallic behavior for $x>0.3$~\cite{Fukazawa:2002hm,Zhu:2014hoa}, the nature of this metallic state remains to be clarified.
It would be interesting to explore the optical conductivity or $T$-dependent resistivity of these doped compounds to confirm the NFL behavior.
A comparative study with the osmate Cd$_2$Os$_2$O$_7$ with a $5d^3$ configuration would also be interesting~\cite{Shinaoka:2012ja,Yamaura:2012ey,Hiroi:2015kr,Tardif:2015es}.
\begin{acknowledgments}
We thank Lewin Boehnke, Xi Dai, Li Huang, Hugo Strand, Jakub Imriska, Yukitoshi Motome, Hongbin Zhang, Youhei Yamaji, Lei Wang, Hiroshi Watanabe for stimulating discussions and useful comments.
We thank Jakub Imriska for the use of his code.
We also thank Kentarou Ueda and Jun Fujioka for dicussions on their experimental data and Shoji Ishibashi for providing the pseudo potentials.
We acknowledge support from the DFG via FOR 1346, the SNF Grant 200021E-149122, ERC Advanced Grant SIMCOFE and NCCR MARVEL.
The calculations have been performed on the M\"{o}nch and Brutus clusters of ETH Z\"{u}rich using codes based on ALPS~\cite{Bauer:2011tz}.
The crystal structure was visualized using VESTA 3~\cite{Momma:2011dd}.
\end{acknowledgments}

\bibliography{ref,ref2}

\appendix
\section{Solving the impurity problem}
In this section we explain how we solved the impurity problem using the hybridization expansion continuous-time quantum Monte Carlo algorithm (CT-HYB).\cite{Werner:2006ko,Werner:2006iz}
There are four iridium atoms in a unit cell.
The effective action for the $m$-th iridium atom is given by
\begin{eqnarray}
S &=& S_\mathrm{loc} \nonumber \\
&& + \sum_{ij}^3 \sum_{\sigma \sigma^\prime} \int_0^\beta  \diff \tau \diff \tau^\prime \Delta_{i\sigma, j\sigma^\prime}(\tau^\prime-\tau) c^\dagger_{i\sigma}(\tau^\prime) c_{j\sigma^\prime}(\tau),\hspace{2mm}~\label{eq:action-cthyb}
\end{eqnarray}
where $\Delta$ is the hybridization function [$\Delta_{i\sigma, j\sigma^\prime}(\tau) = \Delta_{j\sigma^\prime, i\sigma}^*(\tau)$].
$\Delta(\tau)$ is a $6\times 6$ matrix with non-zero off-diagonal elements.
$i$ and $j$ are the index of the $t_\mathrm{2g}$-like Wannier functions,
while $\sigma$ and $\sigma^\prime$ denote spin.
The Wanniner functions are constructed in terms of the local cubic frame of an IrO$_6$ octahedron.
We denote the inverse temperature by $\beta$.

The local Hamiltonian consists of the one-body part $\mathcal{H}_0$ and the Slater-Kanamori interaction $\mathcal{H}_\mathrm{int}$.
The one-body part is given by
\begin{eqnarray}
   \mathcal{H}_0 &=& \hat{P}_m \left(\frac{1}{N_k} \sum_k \mathcal{H}(k) \right)\hat{P}_m,
\end{eqnarray}
where $N_k$ is the number of $k$ points.
The 24$\times$24 matrix $\mathcal{H}(k)$ is the LDA one-body Hamiltonian in the Wannier basis.
$\hat{P}_m$ is a projector to the $m$-th iridium atom.
The Slater-Kanamori interaction $\mathcal{H}_\mathrm{int}$ is defined in terms of the $t_\mathrm{2g}$-like Wannier functions.
Its explicit form is given in Eq.~(1) of the main text.

To avoid a severe sign problem without further approximation,
we transform the single-particle basis as
\begin{eqnarray}
  d_a &=& \sum_b U_{ba}^* c_b,\\
  d^\dagger_a &=& \sum_b U_{ba} c^\dagger_b,
\end{eqnarray}
where $U_{ab}$ is a unitary matrix with $a$ and $b$ being the combined index of spin and orbital.
In practice, we build the matrix $\boldsymbol{U}$ from the eigenvectors of 
\begin{eqnarray}
\mathcal{H}_0 + \delta\times\hat{j}_\mathrm{eff}^{111},~\label{eq:basis}
\end{eqnarray}
where $\delta$ is a positive but very small constant and $\hat{j}_\mathrm{eff}^{111}$ is the effective angular momentum projected on the local [111] axis.

The second term in Eq.~(\ref{eq:action-cthyb}) reads
\begin{eqnarray}
S_\mathrm{hyb} &=& \sum_{\alpha\beta} \int_0^\beta  \diff \tau \diff \tau^\prime \bar{\Delta}_{ab}(\tau^\prime-\tau) d^\dagger_a(\tau^\prime) d_b(\tau),\hspace{3mm}\label{eq:action-cthyb-rot}
\end{eqnarray}
with
\begin{eqnarray}
\bar{\Delta}_{ab} (\tau) &=& \sum_{cd} (U^\dagger)_{ac} \Delta_{cd}(\tau) U_{db}.
\end{eqnarray}

Now, we expand the partition function $Z$ as
\begin{eqnarray}
Z &=& Z_\mathrm{bath}  \sum_{n=0}^\infty \frac{1}{n!^2}\sum_{\alpha_1,\cdots,\alpha_n}\sum_{\alpha^\prime_1,\cdots,\alpha^\prime_n}\nonumber \\
&&  \int_0^\beta \mathrm{d} \tau_1 \mathrm{d} \tau_1^\prime \cdots \int_0^\beta \mathrm{d} \tau_n \mathrm{d} \tau_n^\prime \nonumber\\
&& \mathrm{Tr_{loc}}\left[ e^{-\beta\mathcal{H}_\mathrm{loc}} T d_{\alpha_n}(\tau_n) d^\dagger_{\alpha_n^\prime}(\tau_n^\prime) \cdots d_{\alpha_1}(\tau_1) d^\dagger_{\alpha_1^\prime}(\tau_1^\prime)\right]\nonumber\\
&& \times \mathrm{det} \boldsymbol{\bar{M}}^{-1},\label{eq:z-cthyb}
\end{eqnarray}
where $Z_\mathrm{bath}$ is the partition function of the bath.
The matrix elements of $\boldsymbol{\bar{M}}^{-1}$ at $(i,j)$ are given by the hybridization function $\bar{\Delta}_{\alpha_i^\prime,\alpha_j}(\tau_i^\prime - \tau_j)$.

We perform importance sampling with respect to the partition function using the weight
\begin{eqnarray}
&&w(d_{\alpha_1}(\tau_1), \cdots, d_{\alpha_n}(\tau_n); d_{\alpha_1^\prime}(\tau_1^\prime), \cdots, 
d_{\alpha_1^\prime}(\tau_n^\prime)) \nonumber \\
&=& \Big| \frac{d\tau^n}{n!^{2n}} \mathrm{Tr_{loc}}\Big[ e^{-\beta\mathcal{H}_\mathrm{loc}} T d_{\alpha_n}(\tau_n) d^\dagger_{\alpha_n^\prime}(\tau_n^\prime) \cdots \nonumber \\  && d_{\alpha_1}(\tau_1) d^\dagger_{\alpha_1^\prime}(\tau_1^\prime)\Big] \mathrm{det} \boldsymbol{\bar{M}}^{-1}\Big|.\label{eq:weight}
\end{eqnarray}
The local trace is evaluated using the matrix formalism: $e^{-\tau\mathcal{H}_\mathrm{loc}}$, $d_\alpha$, $d_\alpha^\dagger$ in Eq.~(\ref{eq:weight}) are represented in the eigenbasis of $\mathcal{H}_0$.~\cite{Werner:2006iz}
Note that we do not have to transform $\mathcal{H}_0$ to the new single-particle basis in terms of $d$ and $d^\dagger$.

\section{Effect of the off-diagonal hybridization functions}
\begin{figure}
     \centering
     \includegraphics[width=0.4\textwidth,clip]{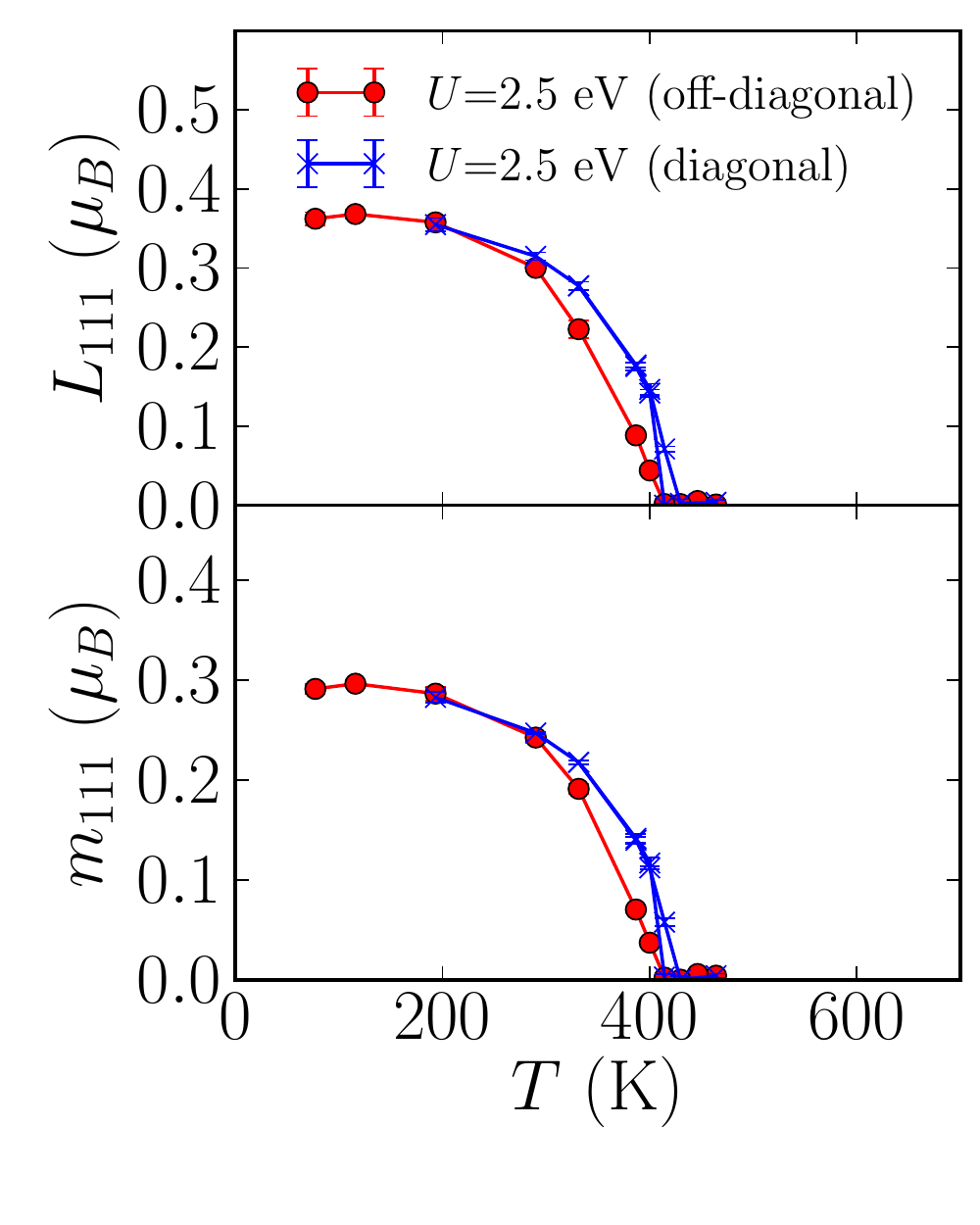}
      \caption{(color online).
      Comparison of the spin and orbital moments computed with/without off-diagonal elements of the hybridization function. A hysteresis behavior is seen in the results computed with the diagonal hybridization function.
    }
 \label{fig:hyb-diag}
\end{figure}

We briefly discuss the effects of the off-diagonal hybridizations in the LDA+DMFT results.
Figure~\ref{fig:hyb-diag} compares the spin and orbital moments computed with/without off-diagonal elements at $U=2.5$ eV.
The behavior near the phase transition is affected by dropping the off-diagonal elements.
In the calculation with diagonal hybridization functions,
the transition is weakly first order, as evidenced by the small hysteretic region.
On the other hand, we observed a continuous transition in the simulation with off-diagonal hybridization functions.

\end{document}